\documentclass[traditabstract]{aa}
\usepackage{graphicx}
\usepackage{epsfig}
\usepackage{color}
\usepackage{txfonts}
\usepackage{natbib}
\bibpunct{(}{)}{;}{a}{}{,} 

\begin{document}

\title{The slowing down of galaxy disks in dissipationless minor mergers}

\titlerunning{Slowing down of galaxy disks in dissipationless minor
mergers}

\author{Yan Qu\inst{1}, Paola Di Matteo\inst{1}, Matthew
  Lehnert\inst{1}, Wim van Driel\inst{1}, Chanda J. Jog\inst{2}}

\authorrunning{Qu et al.}

\institute{GEPI, Observatoire de Paris, CNRS, Universit\'e
  Paris Diderot, 5 place Jules Janssen, 92190 Meudon, France\\
\email{yan.qu@obspm.fr}
\and
  Department of Physics, Indian Institute of Science, Bangalore
  560012, India\\
}

\date{Received, Accepted}

\abstract{We have investigated the impact of dissipationless minor
  galaxy mergers on the angular momentum of the remnant. Our
  simulations cover a range of initial orbital characteristics and the
  system consists of a massive galaxy with a bulge and disk merging
  with a much less massive (one-tenth or one-twentieth) gasless
  companion which has a variety of morphologies (disk- or
  elliptical-like) and central baryonic mass concentrations. During
  the process of merging, the orbital angular momentum is
  redistributed into the internal angular momentum of the final
  system; the internal angular momentum of the primary galaxy can
  increase or decrease depending on the relative orientation of the
  orbital spin vectors (direct or retrograde), while the initially
  non-rotating dark matter halo always gains angular momentum. The
  specific angular momentum of the stellar component always decreases
  independent of the orbital parameters or morphology of the
  satellite, the decrease in the rotation velocity of the primary galaxy
  is accompanied by a change in the anisotropy of the orbits, and the
  ratio of rotation speed to velocity dispersion of the merger remnant
  is lower than the initial value, not only due to an increase in the
  dispersion but also to the slowing -down of the disk rotation. We
  briefly discuss several astrophysical implications of these results,
  suggesting that minor mergers do not cause a ``random walk'' process
  of the angular momentum of the stellar disk component of galaxies,
  but rather a steady decrease. Minor mergers may play a role in
  producing the large scatter observed in the Tully-Fisher relation
  for S0 galaxies, as well as in the increase of the velocity
  dispersion and the decrease in $v/\sigma$ at large radii as observed
  in S0 galaxies.}


\maketitle

\section{Introduction}

Numerical simulations, as well as observations, show that the final
product of a merger between two disk galaxies depends mostly on their
mass ratio. Major mergers of spiral galaxies, i.e. of pairs with mass
ratios ranging from 1:1 to 3:1, have been known for decades to form
pressure-supported elliptical galaxies \citep{barnhern91, barn92,
    bekki197, naabburk199, bendobarn00, springel00, cretton01,
    naabburk03, bournetal05}, with ``boxy'' or ``disky'' shaped
isophotes. Intermediate mass ratio mergers , with mass ratios from
4.5:1 to 10:1, produce hybrid systems which typically have spiral-like
morphologies but elliptical-like kinematics \citep{jog02, bournetal04,
  bournetal05}, while minor mergers (mass ratio $\leq$ 10:1) are much
less violent dynamical processes that do not destroy the disk of the
more massive galaxy in the merger. However, understanding the impact
of minor mergers on galaxies is of particular interest, considering
that they occur frequently in the local universe \citep{frenketal88,
  carlcouch89, laceycole93, gao04, jogeeetal09, kavetal09} and that
they can induce a number of morphological signatures in the final
remnant \citep{ibataetal94, ibataetal01, mart01, new02, yanny03,
  erwetal05, ibataetal05, pohtru06, youngetal07, feldetal08, mart08,
  youngetal08, kazetal09}.

The importance of minor mergers in shaping the evolution of
 galaxies has been supported by detailed numerical simulations. For
example, they can induce changes that move late-type spiral
galaxies towards earlier Hubble types, and contribute to the
thickening and heating of the stellar disk \citep{quinnetal93,
 walker96, velaz99, fontetal01, bensonetal04, kazetal08}. The disks
of galaxies are generally not fully destroyed by these events, and a
remaining kinematically cold and thin component containing $\sim$15 to
$25\%$ of the initial stellar disk mass can indeed survive, embedded in
a hotter and thicker disk \citep[e.g.,]{villahel08}, whereas the
presence of gas in the disk galaxy can largely prevent the destruction
of its thin stellar disk \citep{mosteretal09}.

However, while the role that minor mergers may play in the mass
assembly, kinematics, and morphologies of present-day galaxies has
been studied extensively both observationally and numerically, much
less is known about the way angular momentum is redistributed during
these events. Cosmological simulations show that there is a
discrepancy between the specific angular momentum of dark matter
halos, which have undergone predominantly minor mergers since
$z\sim3$, and the observed baryonic component of present-day
galaxies. The dark matter halos in cosmological simulations have
spin parameters which are a few times smaller than observed in
disks of late-type bulgeless dwarf galaxies \citep{van01,
donghia04}. The evolution of the specific angular momentum of dark
matter halos as caused by major and minor mergers can be
described as a random walk process, with major mergers contributing
to the spin-up of dark halos, and with minor interactions
mostly slowing them down \citep{vitvi02}.

The consequence of an interaction is to transfer orbital angular
momentum into internal rotation: at the beginning of the interaction,
the most extended components first interact tidally, while the more
tightly bound components feel the effects of the encounter only in the
final phases of the merging process \citep{barn92,
  kennicutt98}. However, little is known about the detailed
  redistribution of the angular momentum between the baryonic and
    dark matter components. \citet{mcmillanathana07} have shown that,
    during major mergers of disk galaxies, the dark halos acquire only
    a small amount of rotation, with the dark component of the final
  remnant always having a stellar rotation speed to velocity
  dispersion, or $v/\sigma$, ratio less than 0.2. \citet{jesseit09}
  have discussed the amount of specific angular momentum found in the
  stellar component of major merger remnants, showing that 1:1 mergers
  cause the destruction of the ordered rotation of the progenitor
  disks, while a portion of the original rotation can be preserved
  during mergers with higher mass ratios. If a dissipative component
  (gas) is present in the progenitor disks, gravitational torques
  exerted on it by the stellar disk can extract a significant fraction
  of its initial angular momentum, causing strong gas inflows into the
  circum-nuclear regions \citep{barnhern96}. Finally, angular momentum
  redistribution also occurs during major mergers between spheroidal
  dissipationless galaxies, leading to a transformation of the
    orbital angular momentum into the internal angular momentum of
    both baryons and dark matter, which can be efficient enough to
    produce fast rotating ($v/\sigma > $ 1) stellar halos, as recently
    shown by \citet{dimatteo09}.

A complete and detailed picture of the angular momentum
redistribution during minor mergers is still lacking, even though
  minor mergers are expected to be much more common than major
mergers \citep{fakhouri08} and therefore obviously play an important
role in galaxy mass assembly. For example, \citet{debattista06}
  and \citet{sellwood206} have suggested that minor mergers could
  transfer angular momentum to the dark halo through the action of a
  stellar bar in the disk.

This paper aims to investigate the evolution of the
  angular momentum in minor mergers and its
redistribution between the stellar and the dark matter components of
the primary galaxy and its satellite. We will investigate this process
through the use of dissipationless N-body simulations of minor (mass
ratio 10:1 and 20:1) mergers between a disk plus bulge primary
galaxy and a satellite having either an elliptical or a disk
morphology. Specifically, we will show that independent of the
morphology of the satellite galaxy or the orbital parameters, minor
mergers always cause a slowing down of the stellar disk, as
well as a loss of specific angular momentum within the disk, and
that this process contributes to moving galaxies towards earlier
Hubble types and more slowly rotating systems \citep{emsell07}.
The decrease of the specific angular momentum of the stellar
component of the disk galaxy is accompanied by an increase in
the specific angular momentum of both dark halos (of the primary
and of the satellite), as well as of the pressure-supported stellar
component that was initially part of the satellite.

\section{Models and initial conditions}\label{model}

We studied the coalescence of a massive S0 galaxy and an
elliptical having a mass ratio 10:1 or 20:1. The massive S0 galaxy
(hereafter called gS0) is composed of a spherical dark matter halo and
a spherical stellar bulge, both of which are not rotating
initially, and represented by Plummer spheres \citep{bt1} with total
masses $M_H$ and $M_B$ and core radii $r_H$ and $r_B$, respectively.
The stellar disk is represented by a Miyamoto-Nagai density profile
\citep{bt1} with mass $M_{*}$, and vertical and radial
scale-lengths of $h_{*}$ and $a_{*}$, respectively (Table~\ref{morphtable}). 
The adopted bulge to total baryonic
  mass ratio of our S0 galaxy model, $B/T$ of 1:5, is based on the
  most recent estimate for the ratios of S0 galaxies \citep{laurik07},
  which are up to a factor 3 lower than earlier estimates
  \citep[e.g.,][]{simien86}. As the work of \citet{laurik07} is based on $K_{s}$
  band (2.2 $\mu$m) data, it represents more realistic values for
  the old stellar component than those by \citet{simien86} based on
  $B$ -band data (the dust extinction being more significant at these
  wavelengths). The fraction of dark to baryonic matter inside
  $R_{50}$, the radius containing half of the baryonic matter, is
  $16\%$, which is in broad agreement with observational estimates
  \citep{williams09}, and the resulting rotation curve
  \citep[see][]{chilietal09b} shows a rapid rise in the center,
  followed by a decline at larger radii, typical of many S0s
  \citep{noord07}.

This paper is part of a larger program \citep{dimatteo09} whose
aim is to study systematically the impact of major and minor mergers
on angular momentum redistribution in galaxies. We analyze here the
simple case of the impact of minor mergers on early-type, gas poor,
S0 galaxies and we plan, in future work, to extend this study to the
whole Hubble sequence, with a variety of gas fractions and bulge to
disk ratios. Besides, starting with such simple simulations allows us
to compare our results with a well defined class of objects in order to
verify the appropriateness of our simulations and to gain insight in
possible evolutionary avenues for relatively simple dynamical systems
like S0s.

The satellite elliptical galaxy, which is either ten (dE0, dE0l, dE0h) or
twenty (sE0) times less massive than the gS0 galaxy, consists of
spherical stellar and dark matter components, both modeled with
Plummer profiles, and both not rotating initially. The density
profile of the satellite galaxy is the same in all simulations, and we
only changed the central density of the baryonic component to study
how changing the density would affect the angular momentum
redistribution during the encounter. We thus considered ``reference''
dwarf E0 (dE0) models, as well as galaxies having a central
volume density $50\%$
higher (dE0h) and lower (dE0l) than the reference dE0 galaxy (see
Table~\ref{morphtable}).

To study the dependence of the results on the morphology of the
satellite galaxy, we ran twelve additional simulations, modeling
mergers of the giant gS0 galaxy
with a disk satellite (hereafter called dS0), whose total mass
is ten times smaller. The dS0 consists of a spherical dark
matter halo, a spherical stellar bulge, both of which are not
rotating, and a rotating stellar disk. The morphological parameters
of the dS0 galaxies are given in Table~\ref{morphtable}. We refer
the reader to \citet{chilietal09b} for a more extensive description
of these initial models.

All the initial models and their individual components were evolved in
isolation for 1 Gyr, before starting the interaction. At this
  time, the disk has reached a stable configuration, as discussed in
  the Appendix. The galaxies were initially placed at a
relative distance of 100 kpc, with a variety of relative velocities,
to simulate different orbits. The orbital initial conditions for the
giant-dwarf interactions are given in Table 9 of
\citet{chilietal09b}, and we report here only on the
orbital parameters of the gS0-sE0 simulations (Table~\ref{orbtable2}).

We chose a reference frame with its origin at the barycenter of the
system and an x-y plane corresponding to the orbital plane.
The spin vector of the gS0 galaxy defines initially an angle
$i_1=33^{\circ}$ or $i_1=60^{\circ}$ with respect to the z-axis (see
Fig.~3 in \citet{chilietal09b}). For all simulations involving a
dS0 galaxy, we have chosen an initial angle between the dS0 spin and
the z-axis of $i_2=130^{\circ}$. The orbital angular momentum
can be parallel (direct orbit) or anti-parallel (retrograde orbit)
to the z-axis of the reference frame.

When referring to specific encounters between the spiral and a
satellite, the nomenclature adopted is a six(or seven)-character
string: the first three characters are the type of the massive galaxy,
always "gS0", followed by a three(or four)-character string for dE0
(average concentration, dwarf elliptical satellite), dE0l (low
concentration dwarf elliptical), dE0h (high concentration dwarf
elliptical), dS0 (dwarf spiral satellite) or sE0 (for the two times
less massive elliptical satellite). This is followed by a string
representing the type of encounter (see first column in
Table~\ref{orbtable2} for the specific types), followed by the
suffix ``dir'' or ``ret'', for direct or retrograde orbits,
respectively. The final two numbers, which are either ``33'' or ``60'',
indicate the initial inclination of the gS0 galaxy with respect to
the orbital plane. For example, the nomenclature "gS0dE0h01dir33"
refers to the encounter between the gS0 galaxy (whose initial
inclination $i_1=33^{\circ}$) and the high-concentration elliptical
satellite dE0h, moving on a direct orbit, whose initial orbital parameters are
those corresponding to id = $01dir$ in Table~9 of
\citet{chilietal09b}.

All simulations (50 in total) were run using the Tree-SPH code
described in \citet{benoit02}. A total of $N_{TOT}$=528,000 particles
was used for all simulations,
distributed between the gS0 galaxy and the satellite (see
Table~\ref{numbers}). We also tested the dependence of the results on
the number of particles used in the simulation, running some
additional simulations with a total of $N_{TOT}$=1,056,000 particles.
A Plummer potential was used to soften gravity on small scales,
with constant softening lengths of $\epsilon=200pc$
(or $\epsilon=170pc$ for the high
resolution simulations) for all particles. The equations of motion
are integrated using a leapfrog algorithm with a fixed time step of
0.5 Myr. With these choices, the relative error in the conservation of
the total energy is close to $10^{-6}$ per time step. Since the work
presented here only investigates simulated galaxies without any gas,
only the part of the code evaluating the gravitational forces acting
on the systems has been used.

\begin{table}
\caption[]{Galaxy parameters for the initial models. The bulge and the
  halo are modeled as Plummer spheres, with characteristic masses
  $M_B$ and $M_H$ and characteristic radii $r_B$ and $r_H$. $M_{*}$
  is the mass of the stellar disk whose vertical and radial scale
  lengths are $h_{*}$ and $a_{*}$, respectively.}
\label{morphtable} \centering
\begin{tabular}{lcccccc}
\hline\hline
 & gS0 & dE0l & dE0 & dE0h & dS0 & sE0 \\
\hline
$M_{B}\ [2.3\times 10^9 M_{\odot}]$ & 10 & 7 & 7 & 7 & 1& 3.5\\
$M_{H}\ [2.3\times 10^9 M_{\odot}]$ & 50 & 3 & 3 & 3 & 5 & 1.5\\
$M_{*}\ [2.3\times 10^9 M_{\odot}]$ & 40 & -- & -- & -- & 4 & --\\
$r_{B}\ [\mathrm{kpc}]$ & 2 & 1.7 & 1.3& 1.1 & 0.6 & 0.9\\
$r_{H}\ [\mathrm{kpc}]$ & 10 & 2.2 & 2.2 & 2.2 & 3.2 & 1.55\\
$a_{*}\ [\mathrm{kpc}]$ & 4 & -- & -- & --& 1.3 & --\\
$h_{*}\ [\mathrm{kpc}]$ & 0.5 & -- & -- & --& 0.16 & --\\
\hline
\end{tabular}
\end{table}

\begin{table}
\caption[]{Orbital parameters for the gS0-sE0 interactions}
\label{orbtable2}
\centering
\begin{tabular}{cccccc}
\hline\hline
       orbit id & $r_{ini}^{\mathrm{a}}$ &  $v_{ini}^{\mathrm{b}}$  & $L^{\mathrm{c}}$  & $E^{\mathrm{d}}$& spin$^{\mathrm{e}}$\\
 & & & & &\\
       \hline
       01dir & 100.& 1.48&29.66&0.& up\\
       01ret & 100.& 1.48&29.66&0.& down\\
       02dir & 100.& 1.52&29.69&0.05& up\\
       02ret & 100.& 1.52&29.69&0.05& down\\
       03dir & 100.& 1.55&29.72&0.1& up\\
       03ret & 100.& 1.55&29.72&0.1& down\\
\hline
\hline
\end{tabular}

\begin{list}{}{}

\item[$^{\mathrm{a}}$] Initial distance between the two galaxies, in kpc.

\item[$^{\mathrm{b}}$] Absolute value of the initial relative velocity,
in units of 100 $km s^{-1}$.

\item[$^{\mathrm{c}}$] $L=\mid\bf{r_{ini}} \times \bf{v_{ini}}\mid $,
in units of $10^2km s^{-1}kpc$.

\item[$^{\mathrm{d}}$]  $E={v_{ini}}^2/2-G(m_1+m_2)/r_{ini}$, with
$m_1=2.3\times10^{11} M_{\odot}$ and $m_2=2.3\times10^{10} M_{\odot}$,
in units of $10^4km^2 s^{-2}$.

\item[$^{\mathrm{e}}$]Orbital spin, if parallel (up) or antiparallel
(down) to the z-axis

\end{list}
\end{table}

\begin{table}
\caption[]{Particle numbers for disk galaxy and satellites}
\label{numbers}
\centering
\begin{tabular}{lccc}
\hline\hline
 & gS0 & dE0l,dE0,dE0h,dS0 & sE0 \\
\hline
$N_{star}$ & 320,000 & 32,000& 16,000\\
$N_{DM}$ &160,000 & 16,000& 8,000\\
\hline
\end{tabular}
\end{table}

\section{Results and discussion}

\subsection{Angular momentum evolution}

\begin{figure*}
\centering
\includegraphics[width=14cm,angle=0]{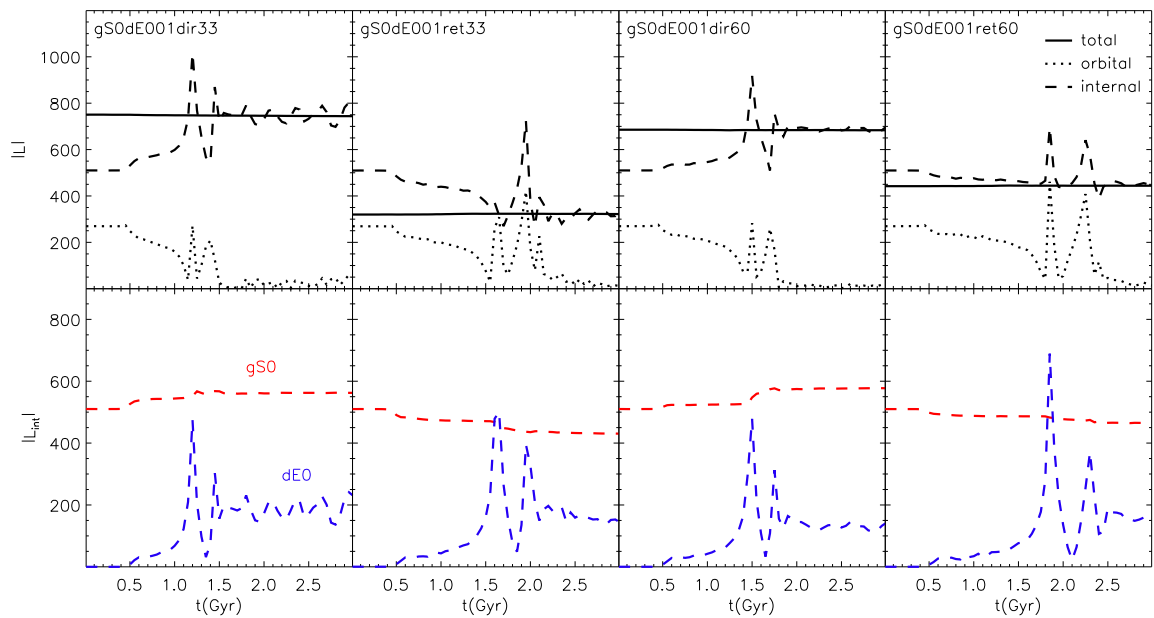}
\vspace{0.8cm}
\caption{Top panels: Evolution of the absolute value of the total
  (solid line), orbital (dotted line) and internal (dashed line)
  angular momentum for some of the simulated 10:1 mergers, for both the
  stellar and dark matter components together. Bottom panels:
  Evolution of the absolute value of the internal angular momentum of the
massive gS0 galaxy (red dashed line) and of the dE0 satellite
  (blue dashed line), for both the stellar and dark matter components
  together. The angular momentum is in units of $2.3\times10^{11}
  M_{\odot}$ kpc km s$^{-1}$.}
\label{AMtot}
\end{figure*}

\begin{figure}
\centering
\includegraphics[width=7cm,angle=0]{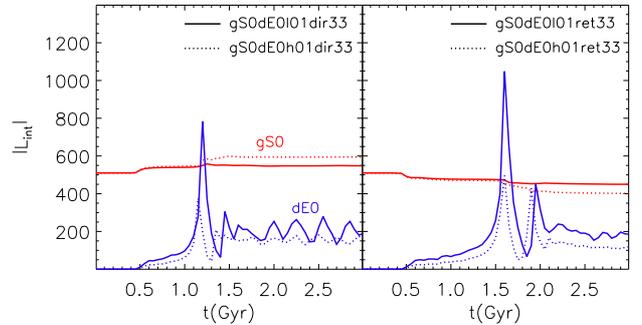}
\vspace{0.8cm}
\caption{Evolution of the absolute value of the internal angular
  momentum in minor mergers with different baryonic mass
  concentrations of the merging satellite. The mass of the satellite
  is the same, only the central volume density is $50\%$ higher
  (dotted lines) or $50\%$ lower (solid lines) than in the reference
  satellite. The internal angular momentum of gS0 galaxy and satellite
  are shown in red and blue, respectively. Left panel: direct
  mergers. Right panel: retrograde mergers.}
\label{AMconc}
\end{figure}

\begin{figure}
\centering
\includegraphics[width=7cm,angle=0]{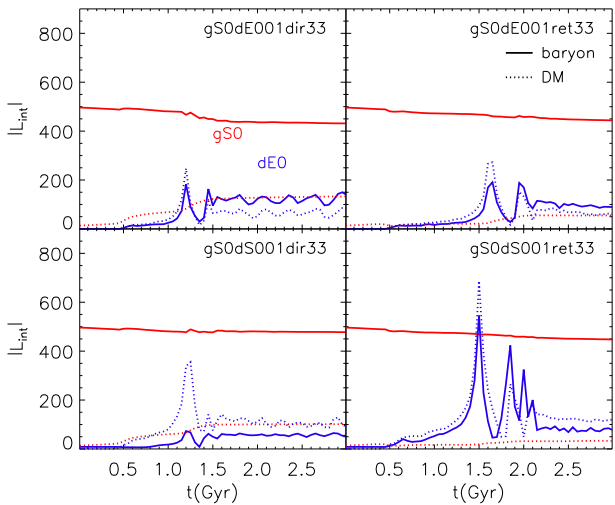}
\vspace{0.8cm}
\caption{Evolution of the absolute value of the internal angular
  momentum of the stellar and dark matter components of the gS0 galaxy
  (solid and dotted red lines) and of its satellite (solid and dotted
  blue lines) for four 10:1 mergers. The angular momentum is in
  units of $2.3\times10^{11}M_{\odot}$ kpc km s$^{-1}$.}
\label{AMcomp}
\end{figure}

As a result of the action of tidal torques and dynamical friction, the
satellite loses its orbital angular momentum (hereafter AM) and
gradually spirals deeper into the gravitational field of the
primary galaxy, where it finally dissolves
completely (meaning it does not survive as an entity within the core
of the primary galaxy). In this section, we discuss how the orbital AM
is converted into internal AM and how it is
redistributed among the different components (stars and dark matter)
of the two galaxies.

\subsubsection{Redistribution of orbital into internal angular momentum}

Fig.~\ref{AMtot} shows the evolution of the absolute value of
the total, orbital and internal angular momenta of the system, for
some of our simulations. In all the cases, the total (internal +
orbital) AM is conserved during the interaction, with an accuracy
of $\lesssim 10^{-6}$ per time step. The orbital AM is
constant until the first pericenter passage of the satellite. At
this time, dynamical friction and tidal torques act on the system,
converting part of the orbital AM into internal AM. This continues
to occur during each successive passage, until the two galaxies
finally merge, and the orbital AM has been completely
converted into internal rotation of the merged system. Looking at the
repartition of the internal AM between the two galaxies
(Fig.~\ref{AMtot}, bottom panels), we can see that the satellite
galaxy always gains part of the orbital AM, while the gS0
galaxy either gains or loses angular momentum, depending on
the relative orientation of its spin and of the orbital AM (direct
and retrograde orbits, respectively). Note that since in
retrograde encounters the orbital AM is anti-parallel to the z-axis
of the reference frame, their total AM is systematically lower
than that of the corresponding direct encounter.

Changing the baryonic mass concentration of the satellite
(Fig.~\ref{AMconc}) affects the redistribution of the internal AM
between the two galaxies, in the sense that, for a given orbit, the
denser the satellite, the smaller the amount of AM it absorbs and the
larger the variation in the internal AM of the primary galaxy. This
trend is due to the fact that a satellite with higher density decays
more rapidly in the central regions of the primary galaxy, since it is
less susceptible to tidal effects of the primary galaxy, preserves a
higher percentage of bound mass, suffers a stronger gravitational
drag. Since tidal torques are less effective on denser satellites (see
\S 3.2 in Di Matteo et al. 2009), the orbital angular momentum is more
efficiently redistributed to the dark halo of the primary galaxy, the
component which mainly drives its orbital decay through dynamical
friction.

\subsubsection{Internal angular momentum of the stellar and dark matter
components}

Here we investigate how the AM is redistributed in each of the two
galaxies, between the stellar and the dark matter components.

In Fig.~\ref{AMcomp} we show how the distribution of the internal AM
between the baryons and dark matter evolves during the merging
process. Near the time of the first barycentric passage, the two
dark matter halos (primary and satellite) acquire part of the
orbital AM, leading to rotation of a halo that initially was
not rotating. In general, the amount of internal AM absorbed by the
dark matter component of the primary galaxy is greater (by a factor
2-3) for pairs on direct orbits than for those on retrograde ones. For
retrograde orbits, we note that the dark halo of the merger remnant,
rotates in the direction opposite to that of the stellar disk,
because it absorbed part of the orbital AM.

Also the stellar component of the satellite absorbs orbital AM
(Fig.~\ref{AMcomp}). For retrograde orbits, this conversion of orbital
into internal AM is a mechanism for creating a counter-rotating old
stellar component in a merger remnant \citep[as first noted by ][for
  minor mergers involving elliptical galaxies]{kor84,
  balcquinn90}. Note that, contrary to the simulations by
\citet{balcquinn90}, our satellite does not survive as a recognizable
entity to the end of the merging process. Its distribution can be
described as ``thick disk-like", encompassing the remaining disk and
extending out to $\sim$10 kpc from the center of the
remnant. Moreover, less than $4\%$ of the internal stellar AM is
transferred to the remnant's bulge by the end of the merger
process. Minor mergers do not add significant amounts of angular
momentum to the bulges of S0 galaxies.

Although none of the initially pressure supported components were
rotating at the beginning of the simulation, they later acquire part
of the orbital AM and begin to rotate. The case of the halo component
is simple: the more massive and extended, the more AM it
absorbs. However, the evolution of the AM of the stellar disk of the
primary galaxy is quite different. In all simulated orbits,
independent of the orbital parameters and the morphology of the
satellite galaxy (whether pressure- or rotationally supported), the
disk of the primary galaxy always loses AM during the interaction.
Although the decrease of the internal AM of the disk varies depending
on orbital parameters and on the internal structure of the satellite
(see Fig.~\ref{AMcomp}), on average the relative change of the
internal AM is $\Delta$L/L$\sim 10\%$ for mergers with a mass ratio
10:1. We also find that satellites with higher central stellar
densities lead to remnants with a lower internal AM in the stellar
component that was formerly part of the disk of the primary.

\subsection{The slowing-down of rotating disks}

\begin{figure*}
\centering \includegraphics[width=12cm,angle=0]{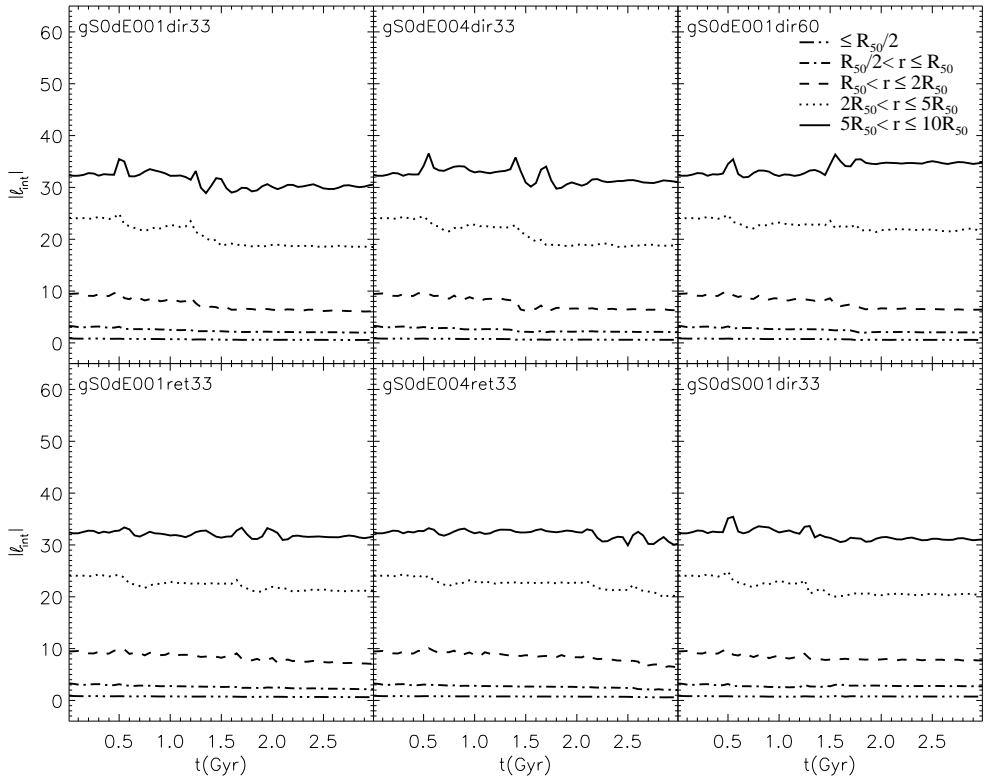}
\vspace{0.8cm}
\caption{Evolution of the absolute value of
   the specific angular momentum of the stars in the
  gS0 galaxy for six of the direct and
  retrograde merger simulations with a dE0 or a dS0 satellite. The
  angular momentum is estimated in five different radial regions
  relative to the half mass radius, $R_{50}$: $r\le0.5R_{50}$,
  $0.5R_{50}<r\le R_{50}$, $R_{50}<r\le2R_{50}$, $2R_{50}<r\le5R_{50}$
  and $5R_{50}<r\le10R_{50}$. The specific angular momentum is in
  units of 100 kpc km s$^{-1}$.}
\label{AMdis}
\end{figure*}

\begin{figure*}
\centering
\includegraphics[width=12cm,angle=0]{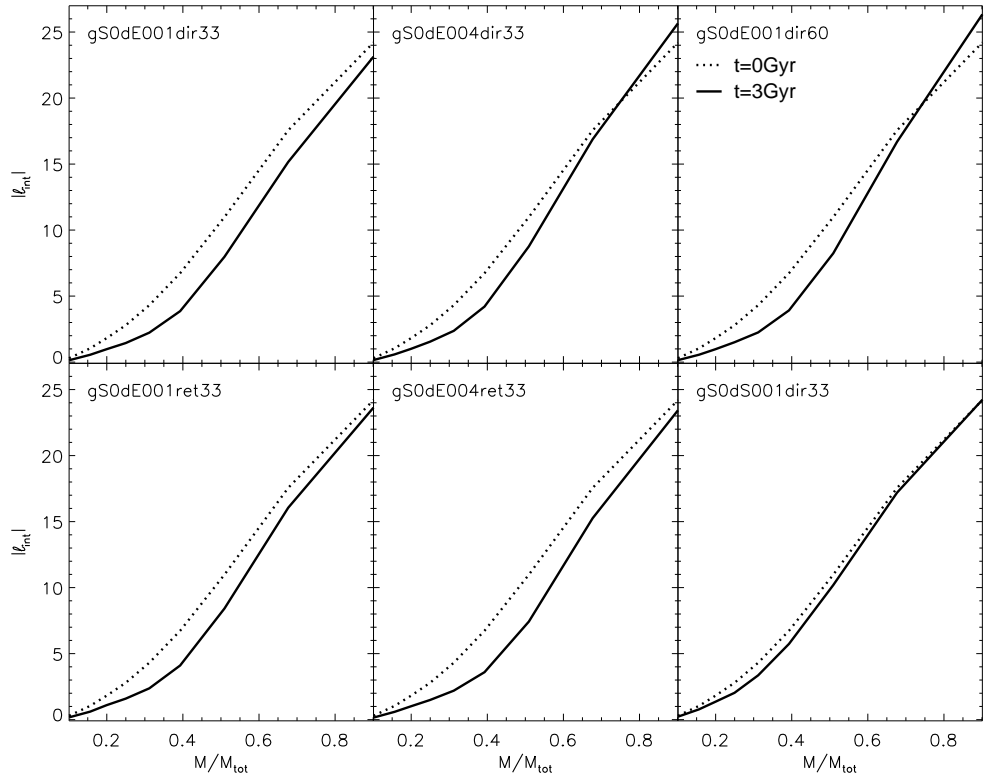}
\vspace{0.8cm}
\caption{Absolute value of the specific
  angular momentum, $l$, of stars in the gS0 galaxy at the start of the
  simulation, t=0 (dotted lines), and at 3 Gyr (solid lines) after the
  start of the simulation, as a function of radii containing a fixed
  percentage of the stellar mass. $l$ was measured at least 1 Gyr after
  the merger completed. The specific angular momentum is in units of
  100 kpc km s$^{-1}$.}
\label{AMmass}
\end{figure*}

\begin{figure*}
\centering
\includegraphics[width=12cm,angle=0]{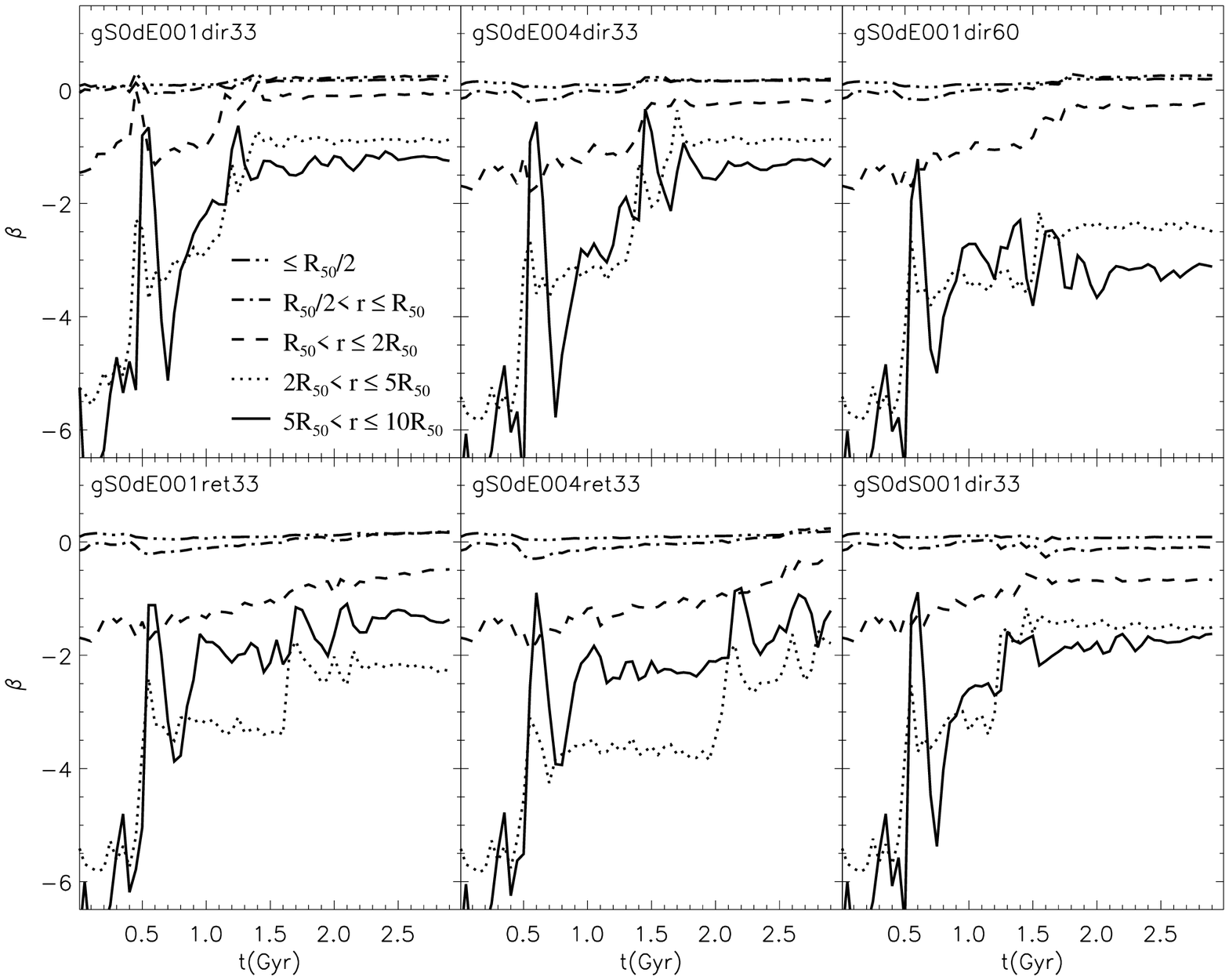}
\vspace{0.8cm}
\caption{Evolution of the anisotropy parameter, $\beta$, of the stars
  in the gS0 galaxy for the six merger simulations whose
   specific internal AM evolutions have been shown in Fig.~\ref{AMdis}
   direct and retrograde merger simulations with a dE0
  or a dS0 satellite. The anisotropy parameter has been measured in
  five different radial regions relative
  to the half mass radius: $r\le0.5R_{50}$, $0.5R_{50}<r\le R_{50}$,
  $R_{50}<r\le2R_{50}$, $2R_{50}<r\le5R_{50}$ and
  $5R_{50}<r\le10R_{50}$.}
\label{AMbeta}
\end{figure*}

The decrease of the internal AM of the primary stellar disk
discussed in the previous section affects the whole of the disk, up
to distances of about 5 to 10 times
$R_{50}$\footnote{$R_{50}=3.3 kpc$ is the initial half-mass radius
of the baryonic component.}. To better demonstrate the loss of
angular momentum, in Fig.~\ref{AMdis} we show the evolution of the
specific internal AM, $l_{int}=<\Sigma_{i} {\bf
r}_{i}\times {\bf
   v}_{i}>$, of the stars
that were initially part of the primary galaxy. We have evaluated
$l_{int}$ for five different radial regions of the galaxy: $r\le
0.5R_{50}$, $0.5R_{50}< r \le R_{50}$, $R_{50}< r \le 2R_{50}$,
$2R_{50}< r \le 5R_{50}$, and $5R_{50}< r \le 10R_{50}$. In all
cases considered here, we find that the stellar component
slows down out to at least $5R_{50}$, i.e. the radius
containing $\sim80\%$ of the total stellar mass (Fig.~\ref{AMmass}),
regardless of the initial orbital parameters, inclination angles, or
satellite morphologies. The same behavior is also found in the
merger with the disky satellite, before it is ultimately destroyed
by the tidal field of the primary galaxy.

The question poses itself what this decrease in rotation speed
is due to? During the merging process, the variation of
the half mass radius, $R_{50}$, is not more than $10\%$, and what
variation there is in the mass distribution is not sufficient to
explain the observed decrease. To gain further insight, we evaluated
the anisotropy parameter, $\beta =
1-\frac{<\sigma^{2}_{t}>}{2<\sigma^{2}_{r}>}$, where $\sigma_{r}$
and $\sigma_{t}$ are the velocity dispersions of the radial
and tangential components, respectively \citep{bt1}. If the velocity
distribution is isotropic then $\beta=0$, and if the velocity
distribution is dominated by tangential motions then $\beta<0$, and
$\beta>0$ if dominated by radial motions. As shown in
Fig.~\ref{AMbeta}, the merging process is accompanied by an
evolution of the $\beta$ parameter in the stellar component. The
change in $\beta$ is particularly large outside of $R_{50}$. In
these outer regions, the stellar orbits, which before the
interaction were dominated by tangential orbits, tend to become
increasingly radially dominated as the merger advances. While the
outermost region, $2R_{50}< r \le 5R_{50}$, of the merger remnant is
still dominated by tangential motions, the motion of stars in the
region $R_{50}< r \le 2R_{50}$ can become nearly isotropic due
to the interaction with the satellite (see for example the
gS0dE001dir33 simulation shown in Fig.\ref{AMbeta}). A similar
change in the anisotropy was previously noted in simulations of single
mergers \citep{bournetal05} and of multiple mergers
\citep{atha05, bournetal07}. In general, the strongest
evolution of the $\beta$ parameter is found in direct encounters.

\begin{figure}
\centering
\includegraphics[width=9.7cm,angle=0]{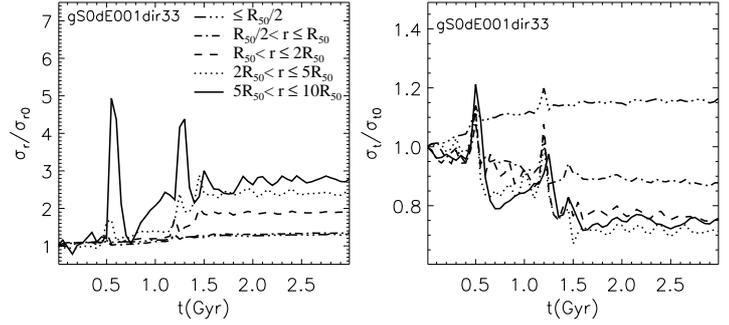}
\caption{The evolution of radial and tangential fractional velocity
  dispersion in different radial regions: $r\le0.5R_{50}$,
  $0.5R_{50}<r\le R_{50}$, $R_{50}<r\le2R_{50}$, $2R_{50}<r\le5R_{50}$
  and $5R_{50}<r\le10R_{50}$.}
\label{AMsigma}
\end{figure}

If the velocity dispersion of a disk increases in a merger, then the
effective rotational velocity decreases, due to the asymmetric
drift, see, e.g., \S 4 of Binney \& Tremaine (1987). The
anisotropy parameter is a way to represent the conversion of
rotational motion into random motions. To confirm that asymmetric
drift is the underlying cause, we evaluated the evolution of the
radial and tangential components of the velocity dispersions in our
simulations. In Fig.\ref{AMsigma}, their fractional evolution (i.e.,
$\sigma_r(t)/\sigma_r(t=0)$ and $\sigma_t(t)/\sigma_t(t=0)$) is
shown for one of our simulations.

The
minor merger is accompanied by an increase of the radial velocity
dispersion at all radii, but particularly in the outermost
regions. At the same time, the transverse velocity dispersion
decreases throughout the whole disk, except in the inner region
(inside $0.5R_{50}$), where the contribution of bulge stars, which
acquire AM during the interaction, leads to an increase of $\sigma_t$.

\subsubsection{The effect of increasing the merger mass ratio to 20:1 }

\begin{figure}
\centering
\includegraphics[width=7.7cm,angle=0]{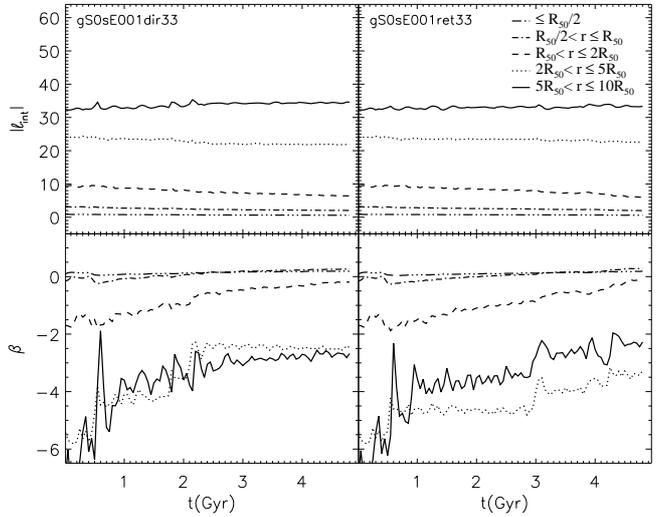}
\vspace{0.8cm}
\caption{Upper panels: Evolution of the absolute value of
   the specific angular momentum of stars in the gS0
  galaxy undergoing a 20:1 direct (left panel) and retrograde (right
  panel) merger. The specific angular momentum has been measured in
  five different radial regions, just as in Fig.~\ref{AMdis}. Lower
  panels: Evolution of the anisotropy parameter, $\beta$, for the
  mergers shown in the upper panels, and measured in the
  same five radial regions.}
\label{newmassratio}
\end{figure}

In the previous sections we have discussed the impact of 10:1 mass ratio
mergers on the redistribution of AM, focusing particularly onto the
cause of the decrease in the rotation speed of the primary stellar
disk. In this section, we want to focus on the effect that
increasing the mass ratio in mergers, specifically from 10:1 to
20:1, has on the evolution of the anisotropy parameter and
the decrease of the specific AM of the disk
(Fig.~\ref{newmassratio}). Also for 20:1 mass ratio mergers
a decrease of the $\beta$ parameter is observed (again
especially in the outer regions), indicating that the velocity
dispersion is increasing in the radial direction. The overall impact
on the disk rotation is, of course, less pronounced than for the
10:1 merger simulations. In this case, we find that the
average decrease in the internal AM of the stellar component is
$\Delta L/L\simeq 8\%$, which is about $30-40\%$ lower than the
value found for 10:1 mergers. However, this is an average
value, whereas the actual decrease in any particular merger depends on
both the morphology and the central density of the satellite galaxy.

\subsection{Rotation speeds and $v/\sigma$}

\begin{figure}
\centering
\includegraphics[width=7.5cm,angle=0]{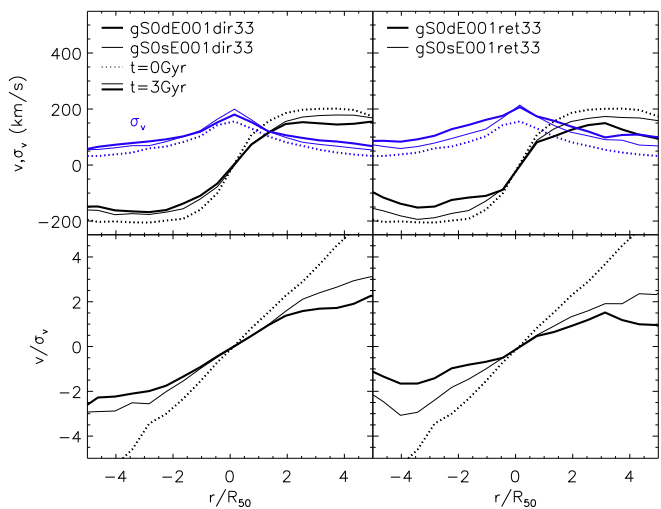}
\vspace{0.8cm}
\caption{Upper panels: Line-of-sight velocities and velocity
  dispersions of stars in the initial gS0 model galaxy (dotted
  lines) and in the final remnant of a 10:1 (thick solid lines) and a
  20:1 (thin solid lines) merger. Direct encounters are shown in the
  left panel, retrograde encounters on the right. Lower panels:
  $v/\sigma$ ratio of the initial gS0 model (dotted lines) and of the
  final remnant of the 10:1 (thick solid lines) and 20:1 (thin solid
  lines) mergers shown in the upper panels. Distances are in multiples
  of $R_{50}$.}
\label{vprofile}
\end{figure}

The decrease in the specific angular momentum of the stellar component
during a minor merger is obviously reflected in other dynamical
properties of the merger remnant. As shown in Fig.~\ref{vprofile} the
minor merger reduces the rotation speed of the stellar component of
the disk galaxy, and increases its velocity dispersion. To
  evaluate these quantities, we considered all stellar particles 1 kpc
  above and below the meridional plane of the primary galaxy,
  including those that were initially part of the satellite galaxy.
In the case of a 10:1 direct merger, the rotation speed at $r=2R_{50}$
decreases by about $25\%$, from the initial value of $v=200km/s$ to
about $v=150km/s$. At the same time the velocity dispersion increases
over the whole disk, e.g. by about $30\%$ at $r=2R_{50}$. This leads
to an overall decrease of the $v/\sigma$ ratio over the entire extent
of the disk, e.g. from 2.3 to 1.4 at $r=2R_{50}$
(Fig.~\ref{vprofile}). The decrease in the rotation speed and
$v/\sigma$ depends on the merger mass ratio: at the end of the 20:1
merger the remnant shows higher values of both $v$ and $v/\sigma$ than
the 10:1 merger. Finally, the decrease in the rotation speed in the
remnants of the two retrograde simulations shown in
Fig.~\ref{vprofile}, gS0dE001ret33 and gS0sE001ret33, is due to both a
slowing down of stars in the gS0 disk (as discussed previously) and to
a negative contribution coming from stars formerly in the satellite
galaxy. These stars have acquired part of the orbital AM, which is in
the direction opposite to the spin of the gS0, and they form a
counter-rotating extended stellar component which contributes to the
overall decrease in the line-of-sight velocity of the remnant galaxy,
and to the increase in its velocity dispersion.

\section{Summary and Discussion}

In this paper, we have studied the redistribution of the angular
momentum during dissipationless minor mergers between a massive
disk galaxy and an elliptical- or S0-like satellite with masses
  that are either ten or twenty times smaller than that of the
primary galaxy. From an analysis of these simulations, we find that:

\begin{itemize}

\item During the merging process the
  orbital angular momentum is redistributed into internal angular
  momentum of both the primary galaxy and the
  satellite;

\item The total internal angular momentum of the primary galaxy can
  either increase or decrease, depending
  on the orientation of the orbital spin (direct or retrograde orbit);

\item While the initially non-rotating dark matter halo of the primary
  galaxy always acquires angular momentum, \emph{the specific angular
    momentum of the stellar component always decreases, independent of
    the initial orbital parameters or the
    morphology of the satellite galaxy};

\item The decrease in the rotation speed of the disk of the main
  galaxy is accompanied by a change in the distribution of the types
  of stellar orbits, especially outside of the half mass radius,
  $R_{50}$. Generally, the radial component of the velocity
  dispersion becomes more and more important, as the merger
  progresses, thus increasing the
  anisotropy parameter, $\beta$, from its
  initially negative value;

\item The ratio of the rotation speed to the velocity dispersion
  of the disk, $v/\sigma$, decreases at all radii due to both an
  increase in the velocity dispersion of the disk, which is
  heated during the merging process, and a decrease in the
  rotation speed of the disk.

\end{itemize}

There are a number of astrophysical implications to the
results we presented of our (rather idealized) simulations,
which we will now discuss very briefly. The Tully-Fisher, or
T-F, relation between the rotation speed and total luminosity
of disk galaxies has, among others, the following two
interpretations. Either the T-F relation originates from the
cosmological equivalence between the dark halo mass and the
circular velocity \citep[e.g.,][]{mo98}, or it is due to
self-regulated star formation in disks of different mass
\citep[e.g.,][]{silk97}. The study of S0 galaxies may help to
constrain which of these hypotheses is correct, and thus to
unravel the evolutionary history of S0 galaxies. Studies of
the differences between the Tully-Fisher relation of S0s and
spirals, while showing some small discrepancies in the offset
and scatter, generally find that the S0 galaxies follow the
spiral relationship, albeit with a larger scatter
\citep[e.g.,][but see
  \cite{mathieu02}]{hinz03, bedregal06}. The offset and larger
scatter have been interpreted as revealing a generally older stellar
population in S0 galaxies than in spirals and a range of
times during which the S0 galaxies started to evolve
passively \citep{mathieu02}. However, as noted in
\citet{bedregal06}, there is no strong
correlation between the age of the S0 galaxies
and the offset from the T-F relation for spiral galaxies. This could be
due to an insufficient sample size, or it
could indicate that other processes may play a role in
the larger scatter in the T-F relation
among S0 galaxies.

We have shown that minor mergers always lead to a decrease in the
rotation speed of the primary galaxy. Decreasing its rotation speed
while increasing its luminosity, will move an S0 galaxy back towards
the spiral T-F relation. While it is premature to attempt to
quantify this effect in the offset, it is clear that
minor mergers would increase the scatter by counteracting, to some
extent, passive evolution. In addition to slowing the rotation speed
of the disk, we also found that the velocity dispersion of the disk
increases, which decreases the v/$\sigma$ of the disk,
especially in the outer parts. As we have show in
Fig.~\ref{vprofile}, this may include an actual decrease with radius
in the very outer parts of the disk. This is qualitatively similar
to what is observed in the dynamics of planetary nebulae in
the outer disks/halos of S0 galaxies \citep{noordermeer08} and
it may suggest that, through simulating multiple minor mergers,
we may be able to reproduce the velocity structure
\citep[e.g.,][]{bournetal07}.

The models of \citet{vitvi02} have shown that, in mergers of
dark matter halos, the specific angular momenta can either increase
or decrease. Over time, this would lead to a increase in the
dispersion of the specific angular momentum of the halos. Our
results on the dark matter halo are consistent with this but we find
systematically a decrease in the specific angular momentum of the
stellar component. So instead of a ``random walk'' over time with only an
increase of the scatter of the angular momentum, our results appear
to suggest a systematic decrease with time in the angular momentum
of the stellar component if minor mergers play an important role in
the growth of galaxies and the redistribution of angular
momentum.

Finally, our results indicate that (single) minor mergers are
capable of moving disk galaxies towards earlier types, since
they reduce the specific angular momentum and rotation speed,
and increase the velocity dispersion of the disk.
\citet{bournetal07} have shown that a sequence of multiple minor
mergers can lead to remnants having global morphological
(flattening, Sersic index, etc.) and kinematical
($v/\sigma$ ratio) similar to those observed in real
elliptical galaxies. Within this context, it will be interesting to
study if the angular momentum content of the stellar and dark
matter components becomes more dispersion dominated and
if size of this effect depends on the total merged mass rather
than on the mass ratio of each merger, and if successive multiple
mergers always slow down the stellar component of the primary
galaxy. Many follow-up questions remain -- for example, is the
specific angular momentum of the stellar component, independent of
the way mass has been assembled in major and multiple-minor
mergers. What role does the dissipative component -- gas -- play in
the way angular momentum is redistributed during the merger and in
the remnant? How does the gas content affect the magnitude of the
decrease in the rotation speed of the stellar disk? Does each minor
merger contribute similarly to the net slowing down, or
does each successive merger become less effective in changing the
angular momentum and rotation speed? Does the net effect saturate, so
that we will ultimately be left with nothing but very slow rotators?

Our simulations show that a single dissipationless minor merger is
sufficient to considerably reduce the $v/\sigma$ ratio, typically by a
factor 1.5-2 at 2$R_{50}$. Our merger remnants can still be classified
as early-type disk galaxies, with a central bulge component and a disk
that is thicker and kinematically hotter than that of the
progenitor. If the cumulative effect of dissipationless multiple minor
mergers leads to systematically slowing down stellar disks, then the
final remnants, which should have an elliptical morphology at that
point, should not exhibit fast rotation at large radii. The existence
of ellipticals with fast rotating halos \citep{rix99, coccato09} may
thus require other formation mechanisms, such as major mergers of two
elliptical progenitors, as recently suggested by \citet{dimatteo09},
or dissipationless mergers of disk galaxies with higher mass ratios,
typically 3:1 according to \citet{bendobarn00}; but see also
\citet{cretton01}. However, the presence of a dissipative component,
whether in the progenitor disks or accreted after the merger, may
change this picture.

We will investigate all these points and their astrophysical implications
as discussed above in subsequent papers.

\section*{Acknowledgments}
YQ and PDM are supported by a grant from the French Agence Nationale
de la Recherche (ANR). PDM thanks the Indo-French Astronomy Network
for a travel grant which made possible a visit to IISc, Bangalore in
August 2009. We are grateful to Beno\^{i}t Semelin and
Fran\c{c}oise Combes for developing the code used in this paper and
for their permission to use it. These simulations will be made
available as part of the GalMer simulation data base
(\emph{http://galmer.obspm.fr}). We wish to thank the referee
for the constructive and helpful report that substantially improved
this manuscript.

\begin{appendix}
\section{Kinematics of the gS0 galaxy when evolved in isolation}

The goal of this paper is to investigate the role minor mergers play
in the angular momentum redistribution and kinematics of the massive
galaxy. Therefore we must distinguish the impact of minor mergers on
the properties of early type disk galaxies from the effects of secular
evolution. To make this distinction, it is also important to study the
evolution of isolated galaxies. To this end, we have run a control
simulation in which the massive S0 galaxy is evolving as an isolated
system over 3 Gyr, i.e., the same duration as the merger runs.

As can be seen in Fig. A1, the kinematics of the isolated S0 galaxy
does not change significantly over this period: no increase is found
in the velocity dispersion of the central regions or the outer disk,
and also the final line-of-sight velocities are remarkably similar to
the initial values, apart from some insignificant decrease outside
2$R_{50}$. Of particular interest is that the final $v/\sigma$ ratio
of the isolated massive galaxy differs significantly from that of the
massive galaxy even after a minor merger, even in the 20:1
case. Comparing Fig.A1 with Fig.\ref{vprofile} shows that the minor
interaction induces a decrease at all radii, even inside $2R_{50}$,
which is not found for the galaxy that evolved in isolation. Secular
evolution causes only a slight decrease in $v/\sigma$ -- the final
ratio is $10\%$ lower than the initial one at r=$R_{50}$, and 23$\%$
at r=$4R_{50}$, whereas even a 20:1 merger induces a much more
pronounced slowing down of the disk at all radii, with a final
decrease in $v/\sigma$ of $30\%$ at r=$R_{50}$ and $42\%$ at
r=$4R_{50}$ (see Table \ref{vsigmatable}). However, it will be of some
interest to find out at what merger mass ratio the decrease in
$v/\sigma$ is comparable to that of secular evolution.

\begin{figure}
\centering
\includegraphics[width=5.cm,angle=0]{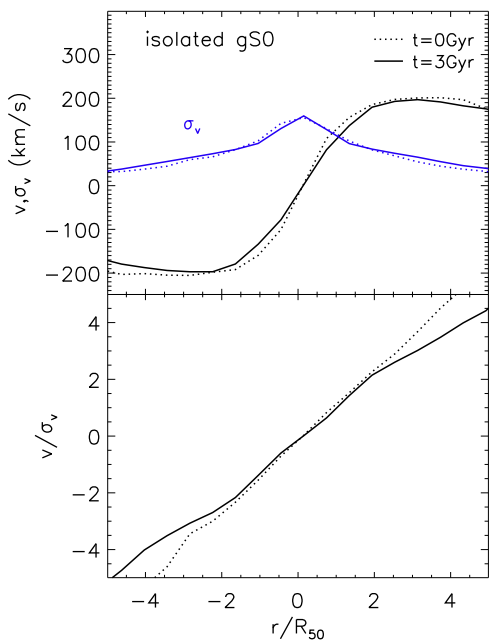}
\vspace{0.8cm}\\
\caption{Upper panel: Line-of-sight velocities and velocity
dispersions of stars in the gS0 galaxy which evolved in isolation, at
t=0 (dotted lines) and at t=3 Gyr (solid lines). Lower
panel: Corresponding $v/\sigma$ ratios at these two times. Distances
are in multiples of $R_{50}$.}
\label{iso-v}
\end{figure}

\begin{table}
\caption[]{$v/\sigma$ ratios at different radii, for the gS0 galaxy which
evolved in isolation and after a minor 20:1 and 10:1 merger. The initial
$v/\sigma$ values are provided in Column 2.}
\label{vsigmatable} \centering
\begin{tabular}{lcccc}
\hline\hline
 & t=0 &  &t=3 Gyr  & \\
 \hline
 &  &   isolated & 20:1 merger & 10:1 merger  \\
 \hline
$v/\sigma$ at $r=R_{50}$ & 1.15 & 1.03&0.8& 0.72\\
$v/\sigma$ at $r=2R_{50}$& 2.44 & 2.14& 1.57& 1.37\\
$v/\sigma$ at $r=4R_{50}$& 4.85 & 3.73& 2.80& 1.81\\
\hline

\hline
\end{tabular}
\end{table}

\end{appendix}
\end{document}